# The external pallidum: think locally, act globally

Connor D. Courtney, Arin Pamukcu, C. Savio Chan

Department of Physiology, Feinberg School of Medicine, Northwestern University, Chicago, IL, USA

Correspondence should be addressed to C. Savio Chan, Department of Physiology, Feinberg School of Medicine, Northwestern University, 303 East Chicago Avenue, Chicago, IL 60611. saviochan@gmail.com

Running title: GPe neuron diversity & function

Abstract (129) + main text (4934) + text box (165): 5228 words

**Acknowledgments**
We thank past and current members of the Chan Lab for their creativity and dedication to our understanding of the pallidum. This work was supported by NIH R01 NS069777 (CSC), R01 MH112768 (CSC), R01 NS097901 (CSC), R01 MH109466 (CSC), R01 NS088528 (CSC), T32 AG020506 (AP), and a Parkinson's Foundation postdoctoral fellowship (CDC). We apologize for not being able to cite earlier studies because of limits on the citation number.

**Abstract** (129 words)

The globus pallidus (GPe) of the basal ganglia has been underappreciated for decades due to poor understanding of its cells and circuits. The advent of molecular tools has sparked a resurgence in interest in the GPe. Here, we review a recent flurry of publications that has unveiled the complexity of the molecular landscape and cellular composition within the GPe. These discoveries have revealed that GPe neurons display a number of circuit features that do not conform to the traditional views of the basal ganglia. Consistent with its broad interconnectivity across the brain, emerging evidence suggests that the GPe plays multifaceted roles in both motor and non-motor functions. Altogether, recent data highlight cellular and functional diversity within the GPe and prompt new proposals for computational rules of the basal ganglia.

**Introduction** (292 words)

Our ability to move is essential to survival. Volitional movements are thought to be controlled by two major forebrain descending pathways—the cortical and basal ganglia circuits. The general organization of the basal ganglia is conserved throughout vertebrate evolution[1]. In the simplified basal ganglia circuit, the cortex provides excitatory inputs to both the striatum (dStr) and the subthalamic nucleus (STN). They in turn project to the substantia nigra pars reticulata (SNr) and internal globus pallidus—the output of the basal ganglia (see **Figure 1**). As simplified models of the basal ganglia emphasize feedforward circuitry, intrinsic feedback circuits are often ignored. Importantly, although most circuit models generally assume a homogenous neuronal population, recent studies argue that neuronal heterogeneity is common across the basal ganglia[2–9].

The external globus pallidus (GPe), as part of the basal ganglia, was once thought to be a simple relay nucleus. Pioneering studies using crude manipulations yielded conflicting results regarding the function of the GPe. However, by targeting specific cell types with improved precision, recent work has revealed that the GPe is more complex than once thought. We now know that the GPe has a heterogeneous neuronal makeup, containing populations with distinct characteristics and functions. Critically, there is now direct evidence that GPe neuron activity plays causal roles in movement[10–13]. These results corroborate neural recordings that show phasic changes in the firing of GPe neurons are associated with body movement[14–16].

Historical reviews of the GPe are available elsewhere[17–19]. Here, we focus on the evolving views of neuron classification in the GPe, with particular emphasis on the recent breakthroughs in the circuit properties and functions of GPe neurons. We review evidence supporting the idea that the unique connectivity of GPe neuron subtypes drives their functions and that these functions go awry in disease.

**Classification of Neurons** (1348 words)

<u>Principal neuron classes</u>

Heterogeneity in the phenotype of GPe neurons was noted in the early 1970's. However, as the descriptions of molecularly-defined GPe neuron subtypes were not established until less than a decade ago[20,21], it was not possible to correlate the cellular features with the molecular identity of neurons. Using advanced tools in rodent models, it is now established that the near-exclusive expression of the calcium-binding protein parvalbumin (PV) or the transcription factor Npas1 identifies the two separate supersets of GABAergic neuron types in the adult GPe[4,12,20–25].

PV-expressing (PV$^+$) neurons and Npas1-expressing (Npas1$^+$) neurons constitute 50% and 30% of the adult GPe, respectively, and are considered as the two principal neuron classes (see **Table 1**). They are distinct across multiple modalities, including axonal patterns, synaptic inputs, electrophysiological properties, alterations in disease states, and behavioral roles[11–13,22,24–28]. The reported abundance of PV$^+$ neurons varies across laboratories in part due to methodological differences (e.g., using cell-specific driver lines, retrograde labeling, fate mapping, and immunohistochemistry)[20–24,26,27,29–32]. Additionally, some of the observed range in the abundance of PV$^+$ neurons is due to the spatial gradients in their distribution. However, by systematically examining a number of transgenic lines, a recent examination definitively confirmed that PV$^+$ neurons account for approximately 50% of neurons in the GPe[25].

Neurons lacking expression of either PV or Npas1 amount to 20% of the GPe. This population includes choline acetyltransferase-expressing (ChAT$^+$) neurons, which amount to 5% of the total GPe neuron population[24,25,27] and project to the cortex[27]. The remaining GPe neurons (~15%) are yet to be rigorously characterized. This neuron population expresses the transcription factor Lhx6 (Lhx6$^+$) but not Sox6 (Sox6$^-$)[25]. Likely due to the relatively low abundance of Lhx6$^+$-Sox6$^-$ neurons, no data are available from the previous single-cell transcriptomic dataset[4,33], and identifying reliable markers for this population of neurons has been challenging. It is important to emphasize that Lhx6 expression should not be considered as a classifier as it is seen across multiple GPe neuron populations, including in PV$^+$, Npas1$^+$, and PV$^-$-Npas1$^-$ neurons. Furthermore, the reported expression of Lhx6 within these subpopulations varies widely across laboratories[22,25,26,31,34] (see **Neuron subclasses**). While it is likely that there are additional rare neuron types yet to be discovered, we now have a near-complete description of the major neuron types within the GPe. Future transcriptome-wide spatial profiling would be useful for the characterization of the rare neuron types in the GPe.

Neuron subclasses

Although PV$^+$ neuron and Npas1$^+$ neuron classes are distinct across multiple modalities (see **Table 1**), such as electrophysiological and synaptic properties, subtle differences exist within each neuron class. Through harnessing molecular tools to target specific GPe neuron populations, recent studies have identified distinct subclasses of PV$^+$ and Npas1$^+$ neurons with unique properties. Within the Npas1$^+$ class, 60% are Foxp2$^+$ (Npas1$^+$-Foxp2$^+$, also known as arkypallidal neurons); these neurons represent the first unique GPe neuron subclass described[35]. Compelling data show the distinct features of Npas1$^+$-Foxp2$^+$ neurons, such as their developmental origin and their electrophysiological, anatomical, and molecular profiles[23]. Importantly, as Foxp2 expression is maintained in adults, the Foxp2-Cre mouse line has been successfully used across labs to capture this population of neurons[13,25,36,37].

The remaining neurons within the Npas1+ class are both Nkx2.1+ and Lhx6+. Unlike Npas1+-Foxp2+ neurons that project exclusively to the dorsal striatum, Npas1+-Nkx2.1+ neurons project to the midbrain, cortex, and reticular thalamus[13,25,27] (see **Synaptic Outputs**). Accordingly, both Npas1+-Foxp2+ and Npas1+-Nkx2.1+ neuron subtypes should be regarded as *bona fide* subclasses, as recent single-cell data[4,33] further confirmed the unique transcriptomic properties of these two Npas1+ neuron subclasses. Notably, these studies led to the discovery that GPe neurons labeled using the Npr3-Cre driver are molecularly and anatomically consistent with Npas1+-Nkx2.1+ neurons[13].

In addition to the two distinct Npas1+ neuron subclasses, single-cell transcriptomic data and immunolabeling suggest the existence of PV+ neuron subpopulations[4]. Multiple lines of evidence point to at least two subtypes of PV+ neurons (see also **Developmental origins**). One of these populations is identified by the marker Kcng4. These neurons share electrophysiological properties, axonal projection patterns, and behavioral roles with canonical PV+ neurons[13]. The most compelling data showing functionally distinct subtypes of PV+ neurons comes from Lilascharoen and colleagues, who identified SNr-projecting and parafascicular thalamus (Pf)-projecting PV+ neurons; these non-overlapping sets of PV+ neurons have distinct synaptic partners, intrinsic properties, and behavioral roles[34]. Converging evidence suggests that Lhx6 expression is weak or absent in the Pf-projecting population[26,34]. However, it remains unclear how the current data regarding PV+ neurons intersect with each other. Therefore, it will be paramount for future studies to reconcile how axonal projection profiles align with molecular marker expression within the PV+ neuron class. This in turn should facilitate our goal to map specific neuron types to functions.

Developmental origins

The literature converges on the idea that both the embryonic origins and the congruent expression of transcription factors govern cell specifications[38]. A large set of transcription factors have been previously identified to govern the formation of the GPe[20,21,39–41]. More recent studies have extended our understanding on the molecular landscape of the GPe and have collectively provided a firmer foundation on the neuronal identification of the GPe[13,25].

The GPe primarily arises from the medial and lateral ganglionic eminence (MGE and LGE)[20,21,23,42] (see **Figure 1**). Immunolabeling for Nkx2.1 is a useful approach for the identification of MGE-derived neurons. However, as only 90% of neurons that arise from the Nkx2.1 lineage maintain Nkx2.1 protein expression in adulthood, this approach is not faithful[25] (see **Evolving classification**). While PV+ neurons are believed to largely arise from the MGE, their expression of MGE-markers varies—over 80% express Nkx2.1 whereas only 25% maintain Lhx6 in adulthood[24,25]. This is consistent with data that suggest the existence of two pools of PV+ neurons that are produced with different temporal patterns and occupy slightly different, but otherwise largely overlapping spatial domains within the GPe[42].

Npas1+ neurons are derived from both the MGE and LGE; MGE-derived Npas1+ neurons are Nkx2.1+ and Lhx6+, while LGE-derived Npas1+ neurons are Foxp2+[20–23,25]. Though developmental origins likely influence transcriptional programs that control the specifications of GPe neurons, Npas1+-Nkx2.1+ neurons and Npas1+-

Foxp2$^+$ neurons have similar electrophysiological properties[25]. In contrast, PV$^+$ neurons and Npas1$^+$ neurons have striking differences in their electrophysiological and synaptic properties (see **Synaptic Inputs**). Taken altogether, these data suggest that neuronal birthplace itself does not dictate the characteristics of GPe neurons.

In addition to the MGE and LGE, a small fraction of GPe neurons are derived from the preoptic area (PoA) and descend from the Dbx1-lineage (Dbx1$^+$)[20,25,40,41,43]. While Dbx1$^+$ neurons that originate from the PoA are known to populate the GPe[20,40], their properties were not fully characterized until recently[25,43]. It is now clear that the Dbx1$^+$ population contains neurons that express, to varying degrees, all established GPe neuron markers[25]. This is consistent with the literature that PoA progenitors give rise to a small, but diverse, contingent set of neurons[40,41]. In particular, most Dbx1$^+$ neurons are Sox6-expressing, and they primarily express PV$^+$ (~70%) and to a lesser extent Npas1$^+$ (~10%). Accordingly, Dbx1$^+$ neurons do not correspond to the Lhx6$^+$-Sox6$^-$ population, which is PV$^-$-Npas1$^-$ (see **Principal neuron classes**). Interestingly, PV$^+$-Dbx1$^+$ neurons display a phenotype that is shared with the general PV$^+$ population, which originates primarily from the MGE[20,21].

By using a progenitor-specific fate map approach, it has been shown that sonic hedgehog (Shh)-expressing progenitors from the PoA give rise to a small fraction of neurons in the GPe. This finding disambiguates the assertion from earlier studies, which conflate cumulative recombinase fate-map from a *bona fide* one[43]. The significance of the robust postmitotic expression of Shh remains to be determined. As the cognate signaling receptor and signaling molecules are enriched in GPe astrocytes[44], it is possible that Shh is critical for cell and circuit assembly as well as the maintenance of cell specifications[45].

**Box 1. Evolving classification of GPe neurons** (165 words)

GPe neurons are often classified into two main groups: 'arkypallidal' and 'prototypic' neurons. The defining features of these neurons have emerged over time since the early 2000s[35,46]. Recent evidence confirmed that arkypallidal neurons directly correspond to the Npas1$^+$-Foxp2$^+$ neuron subclass. These neurons have a unique molecular signature, developmental origin, and axonal projection pattern. On the other hand, prototypic neurons (aka non-arkypalldal neurons) encompass a heterogeneous population of neurons that differ widely in their anatomical projections, electrophysiological properties, and functional output, thus limiting the effectiveness of the classification and contributing to less rigorous comparisons within the field. It is important to note that while all PV$^+$ neurons are prototypic, not all prototypic neurons are PV$^+$. Furthermore, not all GPe neurons with backprojection to the dStr are arkypallidal neurons. We propose the term prototypic be avoided, as it is repeatedly misused, and suggest that GPe neurons be classified using more descriptive designations, such as by molecular signature or principal projection target(s) to avoid confusion.

**Synaptic Inputs** (986 words)

The classic model asserts that the GPe is an intrinsic nucleus within the basal ganglia circuit, receiving inputs exclusively from the dStr and the STN. However, accumulating evidence demonstrates that the GPe receives diverse inputs in addition to these two sources. The recent generation of whole-brain input maps[25,34,47,48] reaffirms that the GPe integrates a wide variety of inputs across a number of systems and should therefore no longer be considered an intrinsic nucleus. The intersection of a complex set of inputs at the GPe allows the transformation of salient cues to motoric responses. Additionally, inputs from non-basal ganglia regions support the proposal that the GPe participates in non-motor functions.

Though different markers and classification schemes are used in the literature, the data from prototypic and arkypallidal neurons coincide with the $PV^+$ vs $Npas1^+$ neuron classification. For simplicity, research findings are discussed in relation to these principal neurons in the following sections. Consistent with the idea that $PV^+$ neurons and $Npas1^+$ neurons receive distinct sets of inputs, the two neuron classes have different electrophysiological properties[13,24,25], suggesting a different mode of input-output transformation between them. The differences in the electrophysiological properties of neuron subclasses are more subtle, and it is still unknown how the complement of inputs varies substantially between neuron subclasses.

GABAergic inputs

The dStr is, by far, the largest input to the GPe—the number of input neurons from the dStr constitutes ~80% of the total projection to the GPe and is at least an order of magnitude larger than that from other brain regions[48,49]. This observation is consistent with the earlier finding that GABAergic synapses amount to over 80% of all synapses in the GPe[17].

Contrary to the simplified basal ganglia model, it has been known since the 1980s that axons from both direct-pathway spiny projection neurons (dSPNs) and indirect-pathway SPNs (iSPNs) converge at the GPe. While iSPN axons terminate exclusively in the GPe, modern techniques confirm that dSPN axons collateralize in both the GPe and the SNr[48–54]. Previous studies showed that dSPNs provide roughly half the number of boutons compared to iSPNs in the GPe. However, more recent estimations of the number of boutons formed by iSPNs is three to eight times higher than that formed by dSPNs[48,49]. Both viral-tracing and patch-clamp studies demonstrate that iSPNs strongly target canonical STN-projecting $PV^+$ neurons. On the contrary, dSPNs appear to target $Npas1^+$ neurons, Pf-projecting $PV^+$ neurons, and $ChAT^+$ neurons within the GPe[27,34,36,37,48,55].

Juxtacellular labeling and intracellular dye-loading of GPe neurons have revealed the presence of local axon collaterals with numerous varicosities, suggesting the presence of lateral GABAergic inhibition within the GPe[13,28,35,56–59]. The existence of local inhibitory networks create an anatomical substrate for functional opponency across neuron subtypes. Although local collaterals are integral to GPe circuit dynamics and downstream network effects[60], they have not been studied in great detail due to the inherent difficulty in identifying and selectively activating individual classes of GPe neurons and their local collateral axons. However, using a combination of transgenic lines and optogenetics, two key findings have recently emerged. First, $PV^+$ neurons provide the largest local input. Second, Npas1$^+$-Foxp2$^+$ neurons do not produce appreciable levels of local connections[13,28,36,37].

Glutamatergic inputs

Recent whole-brain input maps verify that the STN is one of the primary glutamatergic inputs innervating the GPe[34,47,48]. However, the cellular targeting properties of this input was not known until recently. Three independent studies have convincingly demonstrated that STN inputs target PV$^+$ neurons more strongly relative to Npas1$^+$ neurons[12,36,37]. While the neuronal makeup of the STN was generally thought to be homogeneous, recent studies show that STN neurons are more heterogeneous than previously expected[6,7,61–64]. Therefore, it will be important to determine if PV$^+$-targeting neurons within the STN are distinct from those targeting Npas1$^+$ neurons.

Recent studies highlight the existence of cortical inputs to the GPe. Importantly, viral-tracing data argue that the spatially-distributed cortical inputs altogether form the largest source of excitatory inputs to the GPe[25,34,65–68]. Cortical inputs arise primarily from pyramidal tract neurons[7,25,69]. However, the precise downstream targets require further characterization, as data from recent studies are somewhat conflicting. While Karube and colleagues suggest that secondary motor cortex input selectively targets Npas1$^+$-Foxp2$^+$ neurons[65], data from others show that cortical innervations do not target specific neuron subtypes[7,25,34].

In addition to the STN and cortex, the Pf and pedunculopontine nucleus send glutamatergic projections to the GPe[12,34,48,70–72]. These inputs do not show a biased connectivity pattern across PV$^+$ neuron and Npas1$^+$ neuron classes[12].

Neuromodulatory inputs

Viral tracing data show that the dopaminergic input arises from the substantia nigra pars compacta, though we have yet to determine the precise neuronal subpopulation responsible[3,34,47,48]. However, this is likely an underestimation of the total dopamine input; earlier findings suggest that a much broader area, which includes substantia nigra pars compacta, ventral tegmental area, and retrorubral area, gives rise to the dopaminergic innervations in the GPe[61,73–80].

In addition to dopaminergic input, tracing studies reveal inputs from corticotropin-releasing factor-expressing (Crf$^+$) neurons in the paraventricular thalamus and the central amygdala[34,47,48,81]. While the signaling partners of this input remain to be explored, strong evidence argues that PV$^+$ neurons are the primary target of Crf$^+$ input, as they robustly express Crfr1[37,82]. Notably, the central amygdala input forms the second largest extrinsic input targeting the GPe[48], and this projection has been recently implicated in fear learning[81]. Additionally, recent literature emphasizes that PV$^+$ neurons are at the intersection of motor-anxiety circuits[47,81,82]; however, we do not currently know the downstream targets that mediate the anxiety-related response. Altogether, whole-brain connectome analyses suggest a diverse neuromodulatory control at the GPe level[3,34,47,48,83] and reinforce the notion that the GPe is a critical node that interfaces signals from a number of systems.

**Synaptic Outputs** (796 words)

Although major efferent projections from the GPe have been mapped previously[17–19], the circuit elements involved were not previously known. By gaining genetic access to unique neuron types within the GPe, along with other methodological advances, researchers have provided detailed circuit dissections (see **Figure 2**) and functional interrogations of the GPe. These studies collectively reveal that changes in the activity of GPe neurons *in vivo* are not merely passive responses to movement—in other words, GPe neurons have the capacity to tune or maintain ongoing actions.

PV$^+$ neurons

PV$^+$ neurons primarily form the principal inhibitory innervation to the STN[22,24,26,29,30,32]. Consistent with this, optogenetic and chemogenetic stimulation of pan-PV$^+$ neurons or selectively STN-projecting neuron subsets promote movement[10,12,34,84]. In addition, this effect can be recapitulated by direct inhibition of STN neurons[12]. These findings are in agreement with the well-established relationship between STN activity and movement suppression[85–93]. Although earlier studies examined the electrophysiological and anatomical properties of STN-GPe network, the cell-type specificity of the STN input in the GPe was unknown until recently. The latest findings collectively emphasize a closed reciprocal loop formed between the STN and PV$^+$ neurons (see **Synaptic Inputs**), adding critical insight into the cellular constituents involved. The recurrent loop formed between the STN and PV$^+$ neurons has strong implications on the pathological network activity observed in Parkinson's disease (PD) (see **Disease Alterations**). However, it is important to note that optogenetic inhibition of PV$^+$ neurons does not yield any motor responses[10], suggesting the possibility of compensatory network mechanisms, such as the involvement of the local collaterals. Alternatively, the basal ganglia may control motor output by computation through a cascade of logic gates that operate on multiple rules. This idea is supported by the co-activation of dSPNs and iSPNs during volitional movement[94–99].

Additionally, a small subset of PV$^+$ neurons target the Pf (see **Neuron subclasses**). These neurons are not involved in motor control, but are associated with reversal learning[34]. Given that STN-projecting and Pf-projecting PV$^+$ neurons receive distinct sets of synaptic inputs, their respective roles in motor and non-motor functions reinforces the concept that GPe neuron subtypes enact their functions through their unique circuit architecture.

Npas1$^+$-Foxp2$^+$ neurons

Based on relative spike timing, earlier literature inferred that Npas1$^+$-Foxp2$^+$ neurons mediate motor inhibition by broadcasting an inhibitory signal across the dStr[100]—their exclusive projection target. However, this causality had not been demonstrated. Recent optogenetic and chemogenetic approaches demonstrate that stimulation of either pan-Npas1$^+$ neurons or the Npas1$^+$-Foxp2$^+$ subset leads to motor inhibition during both spontaneous ambulation and a trained-task[11–13,36]. Furthermore, optogenetic inhibition of Npas1$^+$ neurons leads to motor promotion[12], arguing that the observed phenotype with excitatory actuators is not a spurious gain-of-function. As discussed earlier, Npas1$^+$-Foxp2$^+$ neurons do not form an extensive local collateral network within the GPe; it is therefore likely that they exert their circuit effects within their terminal field (i.e., within the dStr). While Npas1$^+$-Foxp2$^+$ axons impinge on dendrites of SPNs[11,35], the biological significance of this anatomical arrangement has not been examined.

Npas1$^+$-Nkx2.1$^+$ neurons

In contrast with Npas1$^+$-Foxp2$^+$ neurons, the projection and function of Npas1$^+$-Nkx2.1$^+$ neurons are less well-characterized. These neurons project to the midbrain, cortex, and reticular thalamus[13,25,27]. However, it is unclear if a single Npas1$^+$-Nkx2.1$^+$ neuron projects to multiple brain regions, or if separate neurons are responsible for each of these projections.

The projection to the midbrain is in agreement with previous electrophysiological, neurochemical, and viral tracing data[101–105]. While there is evidence suggesting PV$^+$ neurons target the midbrain[101], additional studies strongly argue that this projection is more consistent with the axonal projection patterns of Lhx6$^+$ neurons[26,106]. The discrepancies may be related to the nuances with the use of rabies virus tracing[107]. While this projection is likely constituted by Npas1$^+$-Nkx2.1$^+$ neurons, it remains to be determined if other neurons target midbrain dopamine neurons. More importantly, the behavioral relevance of this projection awaits further clarification. In particular, the feedback loop formed between GPe and midbrain dopamine neurons (see also **Neuromodulatory inputs**) may create complex circuit effects.

The GPe projections to the cortex arise from Npas1$^+$-Nkx2.1$^+$ neurons and ChAT$^+$ neurons[25,27]. A mixture of neurons in the cortex has been shown to receive this GPe input[27,108,109]. While the functions of these projections await further characterization, recent data suggest that they are involved in regulating sleep[110–112]. Additionally, indirect evidence suggests that Npas1$^+$-Nkx2.1$^+$ neurons may contribute to non-motor functions. Specifically, Nkx2.1$^+$ neurons that are PV$^-$ and Foxp2$^-$ innervate lateral habenula-targeting neurons in the internal globus pallidus; however, it is possible these neurons are Lhx6$^+$-Sox6$^-$ rather than Npas1$^+$. Nonetheless, these data further implicate the contribution of GPe neurons to the limbic aspect of the basal ganglia[2,113].

**Disease Alterations** (1260 words)

The selection of desired actions and concomitant suppression of competing actions is key to volitional movement. Disruption of the ability to facilitate desired movements and inhibit unwanted movements results in disorders such as PD, Huntington's disease (HD), and dystonia. As GPe neurons are critically involved in motor control, we survey the emerging observations of the GPe in these disease conditions.

Parkinson's disease

The loss of nigral dopamine neurons in PD leads to widespread basal ganglia circuit alterations. GPe neuron activity is altered following the loss of dopamine, shifting from autonomous decorrelated pacemaking to highly synchronous burst firing. This aberrant GPe neuron activity correlates with the emergence of motor symptoms in both animal models and human patients[114–117], emphasizing the importance of the GPe in PD. Additionally, one of the hallmarks of PD is the widespread development of synchronized basal ganglia activity known as β-oscillations (13–30 Hz), which are implicated in motor suppression[118–122]. Recent evidence indicates that the GPe plays a critical role in the development of pathological β-oscillations. Specifically, optogenetic inhibition of GPe neurons robustly suppresses β-oscillations in chronic 6-OHDA lesioned rodents, a well-established model of PD. Additionally, optogenetic patterning of GPe neuronal firing at β frequency leads to key circuit changes reminiscent of naturally occurring β-oscillations[123].

Recent studies using animal models of PD show that dopamine neuron loss has cell type-specific effects in the GPe. Notably, following chronic 6-OHDA lesion, Npas1$^+$ neurons display a reduction in spontaneous activity (where fast transmission was intact) *ex vivo*[24]. More recent studies suggest that the hypoactivity of Npas1$^+$ neurons is at least in part due to changes in GABAergic signaling rather than alterations in their intrinsic properties[12]. In contrast, following acute or chronic 6-OHDA administration, PV$^+$ neurons display only subtle or no detectable changes in their firing compared to control mice *ex vivo*[24,34,124].

Synaptic inputs to GPe neurons are altered in models of PD. Following chronic 6-OHDA lesion, STN input targeting PV$^+$ neurons is weakened in *ex vivo* slices. This alteration is likely due to a net loss in synaptic contacts and the associated postsynaptic receptors. In contrast, STN-Npas1$^+$ input is unaltered. These results are in agreement with earlier findings that AMPA and NMDA receptors are downregulated in rodent models of PD[12]. Furthermore, both systemic administration and GPe-specific application of NMDA antagonists are effective in ameliorating motor symptoms in animal models of PD[125–127], arguing that the initial alteration may be useful for compensating the hyperactivity of STN input *in vivo*; however, the loss of NMDA receptor activity is likely maladaptive. The altered glutamate imbalance is further exacerbated by the disrupted glutamate homeostasis that is normally maintained by local astrocytes[128].

In addition to STN input, striatal input to the GPe is also altered in the chronic 6-OHDA model. An *ex vivo* examination of dStr-GPe input revealed a net increase in the strength of the total striatal input to the GPe following chronic 6-OHDA lesion[128]. Although this finding is consistent with the prediction from the classic model, the observed alteration was mediated by a local disruption of glutamate homeostasis within the GPe. However, a more recent examination reveals that the indirect-pathway inputs to PV$^+$ neurons are unchanged. This contrasts with the direct-pathway input to Npas1$^+$ neurons, which is strengthened following chronic 6-OHDA lesion[48]. As the ascending Npas1$^+$ projections to the striatum are strengthened under the same condition[11], it is likely that this recurrent loop contributes to the hypokinetic symptoms of PD. It is unclear at this point why there is a discrepancy between studies; it is worth noting that the method of stimulation (electrical vs. optogenetic) can be a key disparity.

The loss of dopamine neurons alters efferent projections and *in vivo* function of GPe neurons. For example, the GPe-STN projection is strengthened following chronic 6-OHDA lesion[129]. This process is heterosynaptically induced, such that excessive cortical input to STN neurons leads to an increase in the strength of the GPe-STN projection[130], indicating that GPe output can be regulated by wider basal ganglia circuit alterations in pathological states. Importantly, optogenetic activation of pan-PV$^+$ neurons or PV$^+$-SNr neurons rescues motor deficits in both acute[10,34] and chronic[12] 6-OHDA animals. These results strongly argue for the adaptive nature of the GPe outputs. Furthermore, following partial dopamine loss, chemogenetic inhibition of PV$^+$-Pf neurons rescues impairment in a reversal learning task[34], suggesting that PV$^+$-Pf neurons may have therapeutic potential for the cognitive symptoms of PD.

Although optogenetic stimulation of PV$^+$ neurons produces reliable locomotor rescue in 6-OHDA models of PD across multiple laboratories, the duration of the motor improvement following GPe neuron stimulation varies across studies (see **Table 2**). For example, in a chronic 6-OHDA model, the motor improvement following PV$^+$ neuron stimulation only lasts for the duration of the optical stimulation[12]. However, studies using acute 6-OHDA models report motor improvements that persist beyond the duration of the stimulation[10,34,131], suggesting that the recovery period following 6-OHDA lesion may be of critical importance in determining whether motor-promoting effects of PV$^+$ neuron stimulation are acute or persistent. Furthermore, these studies differ in the duration and intensity of their stimulation of PV$^+$ neurons, thus making it difficult to compare results across laboratories. Therefore, it is important for future studies to carefully consider the translational applicability of different 6-OHDA lesion models and stimulation paradigms when interpreting functional results.

Huntington's disease

HD is an inherited neurodegenerative movement disorder that leads to basal ganglia dysfunction. Currently, the precise roles of GPe neurons in HD are still unclear. However, hyperactivity of the GPe is presumed to underlie some of the behavioral symptoms of HD[132,133], and deep brain stimulation of the GPe provides symptomatic relief in human patients[134].

Contrary to the widespread loss of SPNs in the striatum, there is typically no GPe neuron loss in human patients until late in the disease progression[135,136]. Consistent with this observation, GPe neuron loss is only observed at 18 months in the Q175 mouse model of HD. Although the abundance of PV$^+$ neurons is maintained for up to 18 months, the abundance of Npas1$^+$-Foxp2$^+$ neurons is significantly reduced at 18 months, but not at 6 or 12 months[137]. It will be important to determine if the susceptibility of Npas1$^+$-Foxp2$^+$ neurons is associated with pathological hallmarks of HD, such as neuronal intranuclear inclusions and mutant huntingtin load.

Increases in α1, β2/3, and γ2 GABA$_A$ receptor subunits on GPe neurons are observed in human patients[138]. Additionally, increased postsynaptic responses to striatal inputs are seen across the Q175, YAC128, and R6/2 models of HD[139,140]. The data from mouse models argue that increased striatal responses are independent of cell loss in the dStr. Instead, this alteration is likely a homeostatic reaction to decreased iSPN input, as the coupling between cortical input and iSPNs is weakened[141–143]. In return, PV$^+$ neurons are hyperactive. Although the intrinsic properties of PV$^-$ neurons are unaltered, their activity *in vivo* is suppressed by the increased collateral input from PV$^+$ neurons[144].

Dystonia

Dystonia is characterized by uncontrollable muscle contractions, resulting in involuntary and abnormal movements and postures[145–147]. Data from animal models and human patients suggest that abnormal GPe neuron activity contributes to dystonia symptoms[148–152]. Importantly, in the DYT1 mouse model of dystonia, there is reduced firing of PV$^+$ neurons due to impaired function of HCN channels. Pharmacological enhancement of HCN channel activity restores physiological PV$^+$ neuron pacemaking and improves motor performance[153]. These data suggest GPe neurons play important roles in dystonia symptomatology and highlight PV$^+$ neurons as potential therapeutic targets in some forms of dystonia. However, the roles of other neuron subtypes in dystonia models remain to be examined.

**Concluding Remarks** (252 words)

Methodological revolutions have greatly advanced our understanding of the GPe. Its complex neuron types, synaptic inputs, and axonal projection patterns argue that the architecture of GPe does not conform to the traditional direct-indirect model of the basal ganglia. Furthermore, these new data go against the hierarchical organization of the classic model, in which information flows unidirectionally along the input to output of the circuit. In particular, the causal roles that GPe neurons play in movement strongly argue that new theories of how the basal ganglia computes motor commands are needed. For example, can neural signals from the GPe veto decisions that are already committed at the cortical, striatal, or STN level? It is likely that rules governing these outcomes are brain state-dependent; therefore, it will be important to establish the actions of various neuromodulators within the GPe with behavioral context.

We are still in the beginning stages of understanding the full behavioral capacity of the GPe, especially the non-motor aspects. Brain-wide inputs to the GPe have now been mapped; future characterization of physiological properties and activation patterns of these inputs in relation to natural behaviors will help further pinpoint the precise roles of the GPe. This should be a reasonable goal as we make major headways in computer vision techniques for pose estimation. Finally, while a consensus on GPe neuron subtypes and their functions is beginning to emerge, we need to develop more powerful models that consider how they integrate with specific action selection networks both within and beyond the basal ganglia.

**Figure Legends** (562 words)

**Figure 1. Cellular and circuit substrates for motor regulation by the GPe.**

**a**. A simplified circuit diagram showing the key nodes within the basal ganglia. The dorsal striatum and subthalamic nucleus are the two input stations that receive excitatory cortical inputs; the substantia nigra pars reticulata is the output station that provides inhibitory projection to the thalamus. The entire basal ganglia circuit, especially the dorsal striatum, is under the neuromodulatory control of the substantia nigra pars compacta. Size of the target areas (circles) is an artistic rendering based on the volume of the brain areas. The circuit role of the GPe is often ignored.

**b**. A modified circuit model highlighting the central role of the GPe within the basal ganglia. The GPe sends inhibitory projections to all key structures within the circuit.

**c**. Pie chart summarizing the neuronal composition of the mouse GPe. The area of the sectors represents the approximate size of each neuron class. $PV^+$ neurons (which constitute 50% of the GPe) are heterogeneous and assume the canonical role of the GPe by projecting to the subthalamic nucleus and substantia nigra pars reticulata. $Npas1^+$ neurons are 30% of the GPe; they can be subdivided into two subclasses. Consistent with the distinct electrophysiological properties, $PV^+$ neurons and $Npas1^+$ neurons have unique behavioral roles: $PV^+$ neurons are motor-promoting, while $Npas1^+$ neurons are motor-inhibiting. $PV^+$ neuron and $Npas1^+$ neuron classes jointly tune motor behavior through a push-pull mechanism. Local inhibitory connections between these neuron classes may control this process. $ChAT^+$ neurons are ~5% of the total GPe neuron population, show no overlap with other known classes of GPe neurons, and project broadly to cortical areas. $PV^-$-$Npas1^-$ neurons amount to ~15% of the total GPe and are the least characterized neuron subtype.

**d**. Diagram of a cross section of the rodent fetal forebrain illustrating the contributions of lateral ganglionic eminence (LGE), medial ganglionic eminence (MGE), and preoptic area (PoA) to GPe neurons subtypes.

**Figure 2. Efferent projections of GPe neuron subtypes.**

**a**. Map displaying the major projection targets of $PV^+$ neurons and $Npas1^+$ neurons. Projection properties of $ChAT^+$ neurons are summarized in **Table 1**. Note that GPe neuron projections are widespread and extend beyond the traditional basal ganglia nuclei, in contrast to the classic indirect pathway model. Magenta: $Npas1^+$-$Foxp2^+$ neurons; Pink: $Npas1^+$-$Nkx2.1^+$ neurons; Dark green: $PV^+$-$Kcng4^+$ neurons; Light green: $PV^+$-$Lhx6^-$ neurons. Abbreviations: Ctx, cortex; dStr, dorsal striatum; Pf, thalamic parafascicular nucleus; SNr, substantia nigra pars reticulata; SNc, substantia nigra pars compacta; STN, subthalamic nucleus; TRN, thalamic reticular nucleus.

**b**. Schematic of the projection targets of $Npas1^+$ neuron subtypes. $Npas1^+$-$Foxp2^+$ neurons represent 60% of the total $Npas1^+$ neuron class and project exclusively to the dorsal striatum, targeting spiny projection neurons. $Npas1^+$-$Nkx2.1^+$ neurons represent 40% of the total $Npas1^+$ neuron class and project to the cortex, thalamic reticular nucleus, and substantia nigra pars compacta.

**c**. Schematic of the projection targets of identified $PV^+$ neuron subtypes. $PV^+$-$Kcng4^+$ neurons overlap with canonical $PV^+$ neurons; they project to the subthalamic nucleus, substantia nigra pars reticulata and provide collateral inhibition to local GPe neurons. $PV^+$-$Lhx6^-$ neurons project to the thalamic parafascicular nucleus. Size of the target areas (dotted circles) in **b, c** are artistic renderings based on the volume of the target areas and do not reflect the axonal density, a strength, or contacts formed by GPe neuron subclasses.

## Table 1. Properties of GPe neurons

| Neuron class (% of GPe neurons) | Neuron subclass (% of principal class) | Marker expression; (additional feature) | Validated driver and reporter line(s) | [d]Synaptic input(s) | Synaptic output(s) | Neurotransmitter(s) | Functional role |
|---|---|---|---|---|---|---|---|
| PV[+] (50%) | [a]PV[+]-Kcng4[+4,13] | [b]Etv1, Lhx6, Nkx2.1, Npy2r, Scna4b[4,10,13,20–25,29–31,34] | PV-Cre, PV-Flp, PV-tdTomato, [c]Kcng4-Cre[24–26,30,32,34] | STN, dStr (iSPNs)[12,34,36,37,48,49,55] | STN, SNr, GPi[22,24–26,29,30,32,34,59] | GABA | motor promotion[10,12,13,34] |
| | PV[+]-Lhx6[−][26,34] | Nkx2.1, Scna4b[4,34] | PV-Cre, PV-Flp[34] | dStr (dSPNs), Pf, [e]cortex[34] | Pf[26,34] | GABA | reversal learning[34] |
| Npas1[+] (30%) | Npas1[+]-Foxp2[+23] (60%) | Meis2, [b]Penk, Sox6[4,22,23,25,35] | Npas1-Cre, [c]Foxp2-Cre[13,24,25,36,37] | dStr (dSPNs)[37,48] | dStr[11,22–25,35,59] | GABA | motor suppression[1,1–13,36] |
| | Npas1[+]-Nkx2.1[+23,25] (40%) | Lhx6, Npr3, Npy2r, Sox6[4,13,20–25] | Npas1-Cre, [c]Npr3-Cre, Lhx6-GFP[13,25] | dStr (dSPNs)[48] | midbrain, cortex, reticular thalamus[24,25,27] | GABA | [g]motor suppression |
| ChAT[+] (5%) | N/A | Nkx2.1[20,22] | ChAT-Cre[24,27] | dStr (dSPNs and iSPNs)[27] | cortex[25,27] | acetylcholine, GABA[27] | unknown |
| PV[−]-Npas1[−] (15%) | N/A | Lhx6; (Sox6[−])[25] | unknown | unknown | unknown | unknown | unknown |

Footnotes:

[a]Kcng4 demarcates a subset of canonical STN-projecting PV[+] neurons. However, a marker encompassing all STN-projecting PV[+] neurons has yet to be determined.

[b]alternative names: Etv1, ER81; Penk.

[c]subclass-specific driver.

[d]refers to predominant inputs. This is a conservative list; other inputs exist.

[e]other cell types also receive cortical input; however, the cell-type selectivity of these inputs remains to be clarified.

[f]inferred based on retrograde tracing demonstrating preferential dSPN input onto pan-Npas1[+] neurons

[g]inferred based on results with pan-Npas1[+] neuron stimulation and Foxp2[+] neuron stimulation.

Table 2. Behavioral rescue in 6-OHA-lesioned models of Parkinson's disease

| Disease model | Elements stimulated | Approach | Effect | Citation |
|---|---|---|---|---|
| chronic (21+ days post-injection) | GPe neurons | chemogenetic | locomotor rescue for duration of stimulation | 84 |
| chronic (28–42 days post-injection) | $PV^+$ GPe neurons | optogenetic | locomotor rescue for duration of stimulation | 12 |
| acute (3–5 days post-injection) | $PV^+$ GPe neurons | optogenetic | locomotor rescue beyond duration of stimulation | 10 |
| acute (3–10 days post-injection) | $PV^+$-STN GPe neurons | optogenetic | locomotor rescue beyond duration of stimulation | 34 |
| unknown duration | non-specific[a] | electrical | locomotor rescue beyond duration of stimulation | 131 |

Footnotes:
[a] electrodes implanted in the GPi

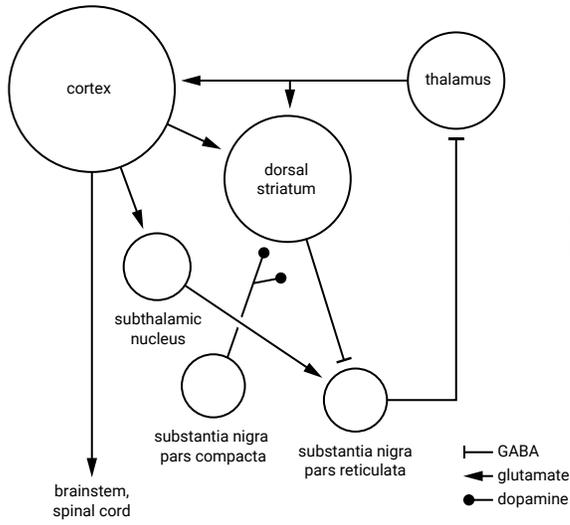
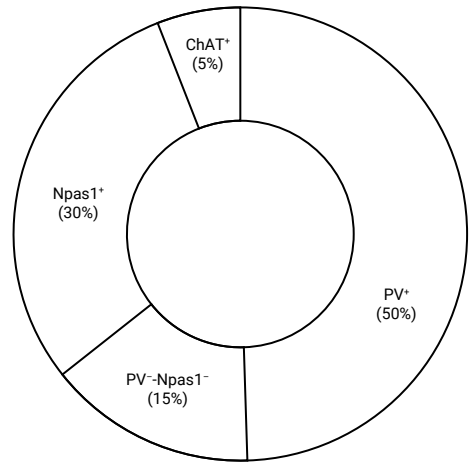
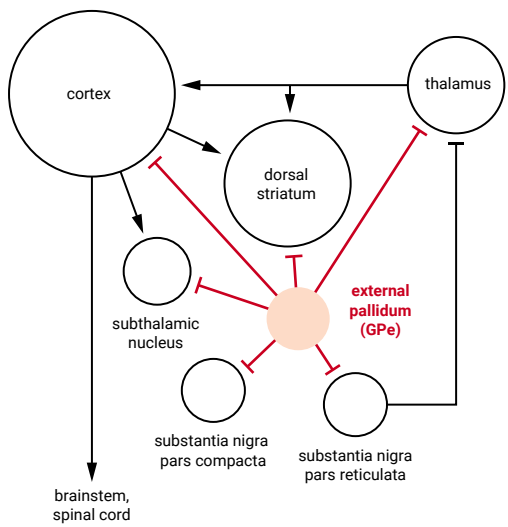
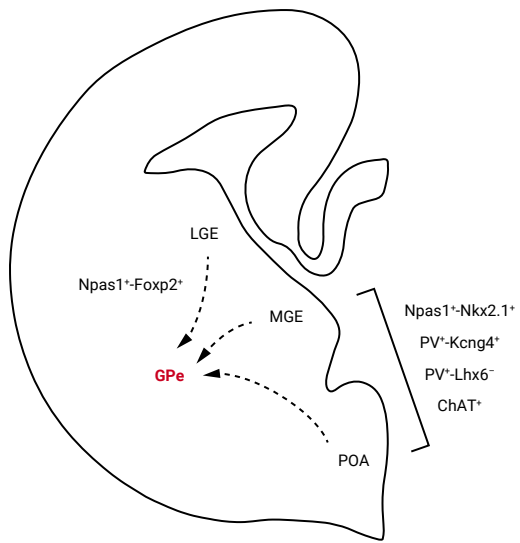

Figure 1

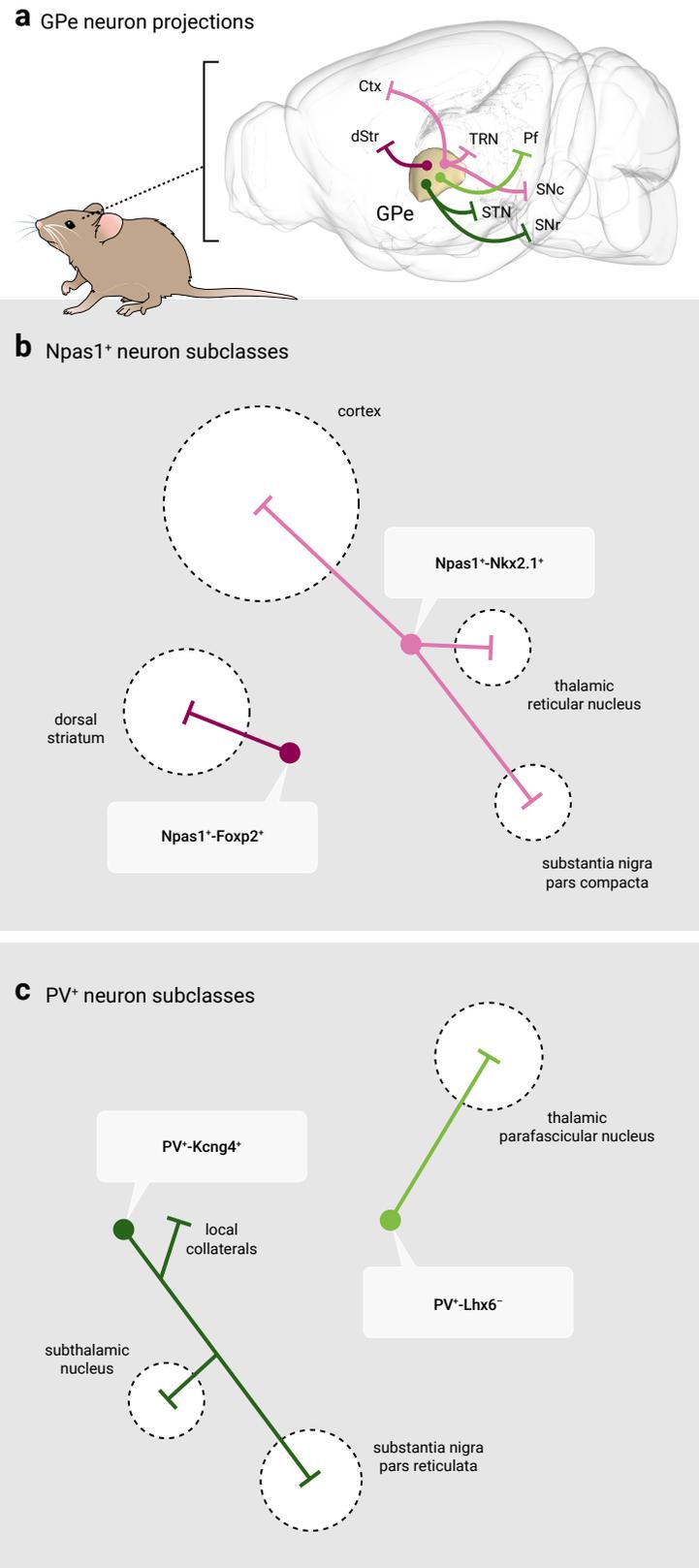

Figure 2